
\magnification 1200
\voffset -1.0 true cm
\hoffset 0.5 true cm
\baselineskip 12pt
\hsize 15.6 true cm
\vsize 25.7 true cm
\phantom{.}
\vskip 4.2 true cm

{\bf \noindent
WEAK MEASUREMENTS}

\vskip 1.3 true cm
\noindent
 \phantom{xxxxxxxxxxx}
Lev Vaidman \hfill \break
\phantom{xxxxxxxxxxx} \hfill \break
\phantom{xxxxxxxxxxx}
 School of Physics and Astronomy\hfill \break
\phantom{xxxxxxxxxxx}
Raymond and Beverly Sackler Faculty of Exact Sciences\hfill \break
\phantom{xxxxxxxxxxx}
 Tel-Aviv University\hfill \break
\phantom{xxxxxxxxxxx} Tel-Aviv, 69978 ISRAEL

\vskip 1. true cm

\noindent
1.  PRE- AND POST-SELECTED QUANTUM SYSTEMS
\vskip .5 true cm

In 1964 Aharonov, Bergmann and Lebowitz$^1$ considered measurements performed
on
a quantum system between two other measurements, results of which were
given. They proposed  describing the quantum system  between two
measurements by using two states: the usual one, evolving towards the future
from the time of the first measurement,
and a second state  evolving backwards in time, from the time of the second
measurement.
 If  a
system  has been prepared at  time $t_1$ in a state $|\Psi_1\rangle$ and  is
found
 at  time $t_2$
in a state $|\Psi_2\rangle$, then at time $t$, $t_1<t<t_2$, the system is
described
by
 $$
\langle \Psi_2 | e^{i\int_{t_2}^{t} H dt}
{\rm ~~~ and~~~}
e^{-i\int_{t_1}^{t} H dt} |\Psi_1\rangle
{}.
 $$
 For simplicity,  we shall consider the free Hamiltonian to
be zero; then the system at time $t$ is described by the two states
 $
\langle \Psi_2 |
$ and
$
|\Psi_1\rangle
$, see Fig. 1. In order to obtain such a system we prepare an ensemble of
systems in
the state $ |\Psi_1\rangle$, perform measurement of desired variable using
separate measuring devices for each system in the ensemble, and perform the
post-selection measurement. If the outcome of the post-selection was not
the desired result, we discard the system and corresponding measuring
device. We look only on measuring devices corresponding to the systems
post-selected in the state $\langle \Psi_2 |$.

 The basic concepts of the  two-state   approach, weak measurement
and  weak values, were developed
several years ago.$^{2,3}$   The weak value of any physical variable $A$ in the
time
interval between pre-selection of the state $| \Psi_1 \rangle$ and
post-selection of the state $ | \Psi_2 \rangle$ is given by
$$
 A_w \equiv {{\langle \Psi_2 | A | \Psi_1 \rangle}
\over {\langle \Psi_2 |\Psi_1 \rangle}} ~~~~.
\eqno(1)$$
  Let us present
the main idea  by way of  a simple example. We  consider, at time $t$, a
quantum system
which was prepared at time $t_1$ in the state $| B = b \rangle$ and was
found at time $t_2$ in the state  $| C = c \rangle$, $t_1 < t <t_2$. The
measurements at times $ t_1$ and $t_2$ are complete measurements of, in
general, noncommuting variables $B$ and $C$.   The  free Hamiltonian is zero,
and therefore,  the first quantum state  at time $t$ is  $| B = b \rangle$.  In
the
two-state  approach we characterize the system at time $t$   by
backwards-evolving state
$\langle C = c| $ as well. Our  motivation for including the future state
is that we know that if a
measurement of $C$ has been performed at time $t$ then the outcome is  $ C = c$
with probability 1. This intermediate measurement,
however, destroys our knowledge that $ B = b$, since the coupling of the
measuring device to the variable $C$ can change  $B$.
The idea of weak measurements is to make the coupling with the measuring device
sufficiently weak so  $B$ does not change.  In fact, we require that both
quantum
states do
not  change, neither the usual
one
 $| B = b \rangle$ evolving towards the future nor $\langle C = c| $
evolving  backwards.

During  the whole time interval
between $t_1$ and $t_2$, both $B  = b$ and  $ C = c $
are  true (in some sense). But  then, $B + C = b + c$ must  also be true. The
latter
statement, however, might not have meaning in the standard quantum formalism
because
the sum of the eigenvalues $b + c$  might not be an eigenvalue of the operator
$ B + C$. An
attempt to measure $B + C$ using a  standard measuring procedure  will lead to
some change of the two quantum states and thus
the outcome will not be  $b + c$. A weak measurement, however, will
yield $b + c$.

 When the ``strong'' value of an observable is known with certainty, i.e.,
we know  the outcome  of an ideal (infinitely strong) measurement with
probability 1, the weak value equal to the strong value. Let us analyze
the example above. The strong value of $B$ is $b$,  its eigenvalue.  The
strong value of $C$ is $c$, as  we know from
{\it retrodiction}.
{}From the definition (1) immediately follows:
$B_w = b$ and  $C_w = c$. But  weak values, unlike  strong values, are
defined   not just for $B$ and $C$ but for {\it all} operators. The strong
value of the sum
$B + C$ when $[B, C] \neq 0$ is not defined, but the weak value of the sum
is: $(B + C)_w = b + c$.
\vskip 1 true cm

\vfill

\noindent
 {\bf
Figure 1.~ Pre- and post-selected quantum system.}~   The
system is considered only if it was found at time $t_2$  in  the state $\vert
\Psi_2 \rangle$ after being  pre-selected at time $t_1$ in the state  $\vert
\Psi_1 \rangle$. The two states yield proper description of the system
for analysis of various measurements at time $t$.

\break

\noindent
 2. QUANTUM MEASURING PROCEDURE
\vskip .5 true cm

In the standard approach to measurements in quantum theory, we measure
observables which correspond to Hermitian operators. The latter have
eigenvalues and a (good) measurement must yield one of these
eigenvalues. If the state of a quantum system is not an eigenstate of the
measured operator, then one can predict only probabilities for different
outcomes of the measurement. The state of the system invariably ``collapses''
to an outcome  corresponding to one eigenvalue. A standard measurement of a
variable $A$ is modeled in the von Neumann
theory of measurement$^4$ by a  Hamiltonian
$$
H = g(t) P A~~~~,
\eqno(2)
$$
where $P$ is a canonical momentum, conjugate to the pointer variable  $Q$ of
the measuring  device.
 The
function  $g(t)$ is nonzero only for a very short time
interval corresponding to the measurement, and is normalized so that
$\int g(t)dt=1$.
During  the time of  this
impulsive measurement, the Hamiltonian (2) dominates the evolution of  the
measured
system and the measuring device. Since $[A , H] =0$, the variable $A$ does
not change during the measuring interaction. The initial state of the
pointer variable is usually modeled by a  Gaussian centered at zero:
$$
\Phi _{in} (Q) =(\Delta ^2 \pi )^{-1/4} e^{ -{{Q ^2} /{2\Delta ^2}}}.\eqno(3)
$$
 Therefore, if the initial state of the system is a superposition
$
|\Psi_1 \rangle = \Sigma \alpha_i |a_i \rangle
$,
then after the interaction (2) the state of the system and the measuring device
is:
$$
(\Delta ^2 \pi )^{-1/4} \Sigma \alpha_i |a_i \rangle e^{ -{{(Q-a_i)^2}
/{2\Delta ^2}}}. \eqno(4)
$$
If   the separation between various eigenvalues $a_i$ is much larger than
the width of the Gaussian $\Delta$, we obtain strict correlation between the
values of the variable $A$ and nearly orthogonal states of the
measuring device. The measuring procedure continues with an amplification
scheme which yields effective (or, according to  some
physicists, real) collapse to one of the pointer positions and the
corresponding  eigenstate  $|a_i \rangle$.
In this  model
the only possible  outcomes of the measurement of the quantum variable $A$ are
the eigenvalues $a_i$. This fact perfectly
matches the premise that the only  values which can be associated with  $A$
are the $a_i$.

When a quantum
system is in a state $|\Psi \rangle$, one can  associate,  mathematically,
to a  variable $A$
 the  the value $\langle \Psi|A|\Psi\rangle$. However,
it was commonly believed that $\langle \Psi|A|\Psi\rangle$ is unmeasurable for
a single system, and
it has physical meaning only for an ensemble of identical systems prepared in
the state $|\Psi \rangle$. For the ensemble, $\langle \Psi|A|\Psi\rangle$ is
interpreted as  statistical average over  the results
 of  measuring  $A$ on this ensemble.
We, however, claim that $\langle \Psi|A|\Psi\rangle$ is more than just a
statistical concept. It has  physical meaning, since it can be measured
directly  on a single system (and not just calculated from the statistics of
the results
$a_i$).

 It
is clear why  standard measurements cannot yield $\langle
\Psi|A|\Psi\rangle$.
The  expectation value is a property of the quantum state and the state itself
is significantly changed during the measuring interaction (2). Thus, in
order to obtain $\langle
\Psi|A|\Psi\rangle$ on a single system we need a  {\it weak} coupling to the
measuring
device. And indeed, under certain conditions, including a  weakened coupling,
von Neumann
 procedure for a measurement of $A$ yields $\langle
\Psi|A|\Psi\rangle$. This procedure is discussed in our  other lecture.$^5$
Here  we want to discuss  weak measurements on
pre- and post-selected quantum systems. The outcomes of such measurements
are the
 weak values defined in Eq. (1).

\vskip 1 true cm

\noindent
3. THE WEAK VALUE IS THE OUTCOME OF WEAK MEASUREMENTS
\vskip .5 true cm

The system  at time $t$ in a pre- and post-selected ensemble is defined by
two states, the usual one evolving from the time of the preparation
and the state  evolving backwards in time from the  post-selection.
We may neglect  the free Hamiltonian if the time between
the pre-selection and  the post-selection is very short.
Consider a system which has been pre-selected in a state $|\Psi_1 \rangle$ and
shortly afterwards post-selected in a state $|\Psi_2
\rangle$.  The weak value of any physical variable $A$ in the time
interval between the pre-selection and the post-selection is given by Eq. (1).
Let us show briefly how weak values emerge from a measuring procedure
with a sufficiently weak interaction.

  We consider a sequence of
measurements: a pre-selection of $|\Psi_{1} \rangle$, a (weak)
measurement interaction of the form of Eq. (2), and a
post-selection measurement finding the state $|\Psi_2 \rangle$.  The
state of the measuring device after this sequence is given (up to
normalization) by
$$
\Phi (Q) = \langle \Psi_2 \vert
e^{-iPA}
\vert \Psi_1 \rangle e^{ -{{Q ^2} /{2\Delta ^2}}} ~~~~.
\eqno(5)
$$
After simple algebraic manipulation we can rewrite it (in the
$P$-representation) as
$$
\displaylines{~\tilde \Phi (P) =   \langle \Psi_2
\vert
 \Psi_1 \rangle ~ e^{-i {A_w} P} ~
e^{-{{\Delta}^2 {P^2}} /{2}} \hfill\cr
\hfill{} + \langle \Psi_2 \vert
 \Psi_1 \rangle  \sum_{n=2}^\infty {{(iP)^n}\over{n!}}
[(A^n)_w - (A_w)^n]   e^{ -{{\Delta ^2 P^2}} /{2}}~.~~ (6)\cr}
$$
\noindent
If $\Delta$ is sufficiently large, then we
can neglect the second term of (6) when we Fourier transform
 back to the  $Q$-representation.  Large $\Delta$
corresponds to weak measurement in the sense that the
interaction Hamiltonian
(2) is small.  Thus, in the limit of weak measurement, the final state
of the measuring device (in the $Q$-representation) is
$$\Phi  (Q) = (\Delta^2 \pi )^{-1/4} e^{ -{{(Q - A_w)^2} /{2\Delta
^2}}}~~~~. \eqno(7)$$
This state represents a measuring device pointing to the weak value, $A_w$.

 Weak  measurements on pre- and post-selected
ensembles yield, instead of eigenvalues, a value which might
lie far outside the range of the eigenvalues.
Although we have showed this result for a specific von Neumann model of
measurements, the result is completely general: any  coupling
of a pre- and post-selected system to a
 variable $A$, provided the coupling
is sufficiently weak, results in effective
coupling to $A_w$. This  weak coupling between  a  single system and the
measuring device
will not, in most cases, lead to a distinguishable shift of the pointer
variable, but collecting the results of  measurements on an  ensemble of
 pre- and post-selected systems will yield the weak values of
a measured variable to any desired precision.

When the strength of the coupling to the measuring device goes to zero, the
outcomes of the measurement invariably yield the weak value. To be more
precise, a measurement  yields the real part of the weak value. Indeed,
the weak value is,
in general,  a complex number, but its imaginary part will contribute only a
phase to the wave function of the measuring device in the position
representation of the pointer. Therefore, the imaginary part will not affect
the probability distribution
of the pointer position which is what we see in a  usual measurement.
However, the imaginary part of the weak value also has physical meaning. It
expresses itself as a change in the conjugate momentum of  the pointer
variable.$^3$

\break

\vskip 1 true cm

\noindent
4. AN EXAMPLE: SPIN MEASUREMENT
\vskip .5 true cm

Let us consider a simple the Stern-Gerlach experiment: measurement of a spin
component of a spin-1/2 particle. We shall consider a particle prepared in the
initial state
spin ``up'' in the $\hat{x}$ direction and post-selected to be ``up'' in the
$\hat{y}$ direction. At the intermediate time we  measure, weakly, the spin
component in the  $\hat{\xi}$ direction which is bisector of  $\hat{x}$ and
$\hat{y}$, i.e.,
$
\sigma_\xi =   (\sigma_x + \sigma_y)/\sqrt 2
$. Thus
${|}\Psi_1 \rangle =|{\uparrow_x} \rangle$,
$|\Psi_2 \rangle =|{\uparrow_y} \rangle$, and  the weak
value of $\sigma_\xi$  in this case is:
$$
(\sigma_\xi)_w =
 {{\langle{\uparrow_y} |\sigma_\xi  |{\uparrow_x} \rangle}
\over {\langle{\uparrow_y} |{\uparrow_x} \rangle}} =
 {1\over\sqrt 2}{{\langle{\uparrow_y} | (\sigma_x + \sigma_y)  |{\uparrow_x}
\rangle}
\over {\langle{\uparrow_y} |{\uparrow_x} \rangle}} = \sqrt 2
 ~~.
\eqno(8)$$
This value is, of course, ``forbidden'' in the standard interpretation
where a spin component can obtain the (eigen)values $\pm1$ only.

An effective Hamiltonian for measuring $\sigma_\xi$ is
$$
H = g(t) P \sigma_\xi~~~~.
\eqno(9)
$$
Writing  the initial state  of the particle in the $\sigma_\xi$
representation, and assuming  the  initial state (3) for  the measuring
device, we obtain that after the measuring interaction   the quantum state of
the system and the pointer of the
 measuring device is
 $$
 \cos {(\pi/8)} |{\uparrow_\xi} \rangle e^{ -{{(Q-1)^2} /{2\Delta ^2}}} + \sin
{(\pi/8)} |{\downarrow_\xi} \rangle  e^{ -{{(Q+1)^2} /{2\Delta ^2}}}~~.
 \eqno(10)
$$
The probability  distribution of the pointer position, if it is observed now
without post-selection,  is the sum of the distributions
for each spin value. It is, up to normalization,
 $$
 prob(Q) = \cos^2{(\pi/8)} e^{ -{{(Q-1)^2} /{\Delta ^2}}} + \sin^2
{(\pi/8)}  e^{ -{{(Q+1)^2} /{\Delta ^2}}}~~.
 \eqno(11)
$$
In  the usual
 strong measurement  $\Delta \ll 1$. In this case, as shown on  Fig. $2a$,
probability distribution  of the pointer is localized around $-1$ and $+1$ and
it is
strongly correlated to the values of the spin, $\sigma_z = \pm1$.

Weak measurement corresponds to   $\Delta$ which is much larger than the range
of the eigenvalues, i.e.,  $\Delta \gg 1$. Fig. $2b$ shows that the pointer
distribution has a  large uncertainty, but it is peaked between the
eigenvalues, more precisely, at the expectation value $\langle{\uparrow_x}
|\sigma_\xi  |{\uparrow_x} \rangle = 1/\sqrt 2$.
An outcome of an individual measurement usually will not be close to
this number, but it can be found from an ensemble of such measurements, see
Fig. $2c$. Note, that we have not yet
considered the post-selection.

In  order to simplify the
analysis of  measurements on  the pre- and post-selected ensemble, let us
assume that we first make the post-selection
of the spin of the particle and only then look on the pointer of the device
that weakly measures $\sigma_\xi$. We must get  the same result as if we
first look at the outcome of the weak measurement,  make the
post-selection, and discard all readings of the weak measurement corresponding
to  the cases in which the result is  not $\sigma_y
=1$. The post-selected state of the particle in the $\sigma_\xi$ representation
is
$
|{\uparrow_y} \rangle =  \cos {(\pi/8)} |{\uparrow_\xi} \rangle -  \sin
{(\pi/8)}|{\downarrow_\xi} \rangle
$.
 The state of the measuring device after the post-selection of the spin
state is obtained by projection of (10) onto the post-selected state:

 $$
 \Phi (Q) ={\cal N} \Bigl(\cos^2 {(\pi/8)} e^{ -{{(Q-1)^2} /{2\Delta ^2}}} -
\sin^2
{(\pi/8)} e^{ -{{(Q+1)^2} /{2\Delta ^2}}})\Bigr)~~,
 \eqno(12)
$$

\break
\phantom{xxxx}
\vfill
\noindent {\bf
Figure 2.~ Spin component measurement without post-selection.}~ Probability
distribution of the pointer variable for measurement of
$\sigma_\xi$ when the particle is pre-selected in the state $\vert
{\uparrow_x} \rangle$.  ($a$) Strong measurement, $\Delta = 0.1.$
($b$) Weak  measurement, $\Delta = 10$.  ($c$) Weak measurement on the ensemble
of 5000 particles. The original width of the peak, 10, is reduced to
$10/\sqrt 5000 \simeq 0.14$. In the strong measurement ($a$) the pointer is
localized
around the eigenvalues $\pm1$, while in the weak measurements ($b$) and ($c$)
the peak is located  in the expectation value $\langle {\uparrow_x}|
\sigma_\xi|{\uparrow_x} \rangle = 1/\sqrt2$.

\break
\noindent
where ${\cal N}$ is a normalization factor. The probability distribution of the
pointer variable is given by
 $$
 prob(Q) ={\cal N}^2 \Bigl (\cos^2 {(\pi/8)} e^{ -{{(Q-1)^2} /{2\Delta ^2}}} -
\sin^2
{(\pi/8)} e^{ -{{(Q+1)^2} /{2\Delta ^2}}})\Bigr)^2~~ .
 \eqno(13)
$$
If the measuring interaction is strong,  $\Delta \ll 1$, then the
distribution is localized around the eigenvalues $\pm 1$ (mostly around 1
since the pre- and post-selected probability to find $\sigma_\xi =1$ is
more than 85\%), see Figs. $3a,3b$. But when the strength of the coupling is
weakened, i.e., $\Delta$ is increased, the  distribution gradually changes to
a single broad peak around $\sqrt 2$, the weak value, see Figs. $3c-3e$.

The width of the peak is large and therefore each individual reading of the
pointer usually will be pretty far from $\sqrt 2$. The physical meaning of
the weak value, in this case, can be associated only with  an ensemble of pre-
and post-selected particles. The accuracy of defining the center of the
distribution goes as $1/\sqrt N$, so increasing $N$, the number of
particles in the ensemble, we can find the weak value with any desired
precision, see  Fig. $3f$.

In our example, the weak value of the spin component is $\sqrt 2$, which is
only slightly more
than the maximal eigenvalue, 1. By appropriate choice of   the pre- and
post-selected states we can get pre- and post-selected ensembles with
arbitrarily large weak value of a spin component. One of the first
proposals$^6$ was to obtain
$(\sigma_\xi)_w = 100$. In this case the post-selected state is nearly
orthogonal to the pre-selected state and, therefore, the probability to
obtain appropriate post-selection becomes very small. While in the case of
$(\sigma_\xi)_w = \sqrt 2$ the  (pre- and)
post-selected ensemble was just half of the pre-selected ensemble, in the
case of $(\sigma_\xi)_w = 100$ the post-selected ensemble   will be smaller
than the original  ensemble by the
factor of $\sim 10^{-4}$.

\vskip 1 true cm

\noindent
5. WEAK MEASUREMENTS ON A SINGLE  SYSTEM
\vskip .5 true cm

We have shown that weak measurements can yield very surprising values which are
far from the range of the eigenvalues. However, the uncertainty of a single
weak measurement (i.e., performed on a single system) in the above example is
   larger than the deviation from the range of the eigenvalues. Each single
measurement separately yields almost no information and  the weak value
arises only from the statistical
average on the ensemble. The weakness and uncertainty of the measurement
goes together. Weak measurement corresponds to small value of $P$ in the
Hamiltonian (2) and, therefore, the  uncertainty in $P$ has to be small. This
requires  large  $\Delta$, the uncertainty of the
pointer variable. Of course,
we can construct measurement with large uncertainty which is not weak at
all, for example, by preparing    the measuring device in a mixed state
instead of a  Gaussian,
 but no precise measurement with
weak coupling is possible. So, usually, a weak measurement on a single
system will not yield the weak value with a good precision. However, there
are special cases when it is not so. Usual strength measurement on a single
pre- and post-selected system can yield ``unusual'' (very different from
the eigenvalues) weak value with a good precision. Good precision means
that the uncertainty is much smaller than the deviation from the range of
the eigenvalues.

Our example above  was not such a case. The weak value
$(\sigma_\xi)_w = \sqrt 2$ is larger than he highest eigenvalue, 1, only by
$\sim 0.4$, while the uncertainty, 1,  is not sufficiently large for
obtaining  the peak of the
distribution near the weak value, see Fig. 3$c$.  Let us modify our
experiment in  such a way  that a single experiment will yield meaningful
surprising result.

\break
\phantom{xxxx}
\vfill
\noindent {\bf
Figure 3.~ Measurement on pre- and post-selected ensemble.}~  Probability
distribution of the pointer variable for measurement of
$\sigma_\xi$ when the particle is pre-selected in the state $\vert
{\uparrow_x} \rangle$ and post-selected in  the state $\vert
{\uparrow_y} \rangle$. The  strength of the
measurement  is
parameterized by the width of the distribution $\Delta$.
 ($a$) $\Delta = 0.1$; ($b$) $\Delta = 0.25$; ($c$) $\Delta =
1$; ($d$) $\Delta = 3$; ($e$) $\Delta = 10$.   ($f$) Weak measurement on the
ensemble
of 5000 particles; the original width of the peak, $\Delta = 10$, is reduced to
$10/\sqrt 5000 \simeq 0.14$. In the strong measurements ($a$)-($b$) the pointer
is localized
around the eigenvalues $\pm1$, while in the weak measurements ($d$)-($f$)
the peak of the distribution is located in the weak value $(\sigma_\xi)_w =
\langle {\uparrow_y}|
\sigma_\xi |{\uparrow_x} \rangle/\langle {\uparrow_y}|{\uparrow_x} \rangle
= \sqrt2$. The outcomes of the weak measurement on the ensemble
of 5000 pre- and post-selected particles, ($f$), are clearly outside the
range of the eigenvalues, (-1,1).

\break

 We consider a system of $N$ spin-1/2 particles all prepared in the state
$|{\uparrow_x} \rangle$ and post-selected in the state $|{\uparrow_y} \rangle$,
i.e., $|\Psi_1\rangle = \prod_{i=1}^N |{\uparrow_x} \rangle _i$ and
$|\Psi_2\rangle = \prod_{i=1}^N |{\uparrow_y} \rangle _i$.
The variable which is measured at the intermediate time is $A
\equiv(\sum_{i=1}^N
(\sigma_i)_\xi)/N$. The operator $A$ has  $N+1$ eigenvalues equally spaced
between $-1$ and $+1$, but the weak value of $A$ is
$$
A_w =
{{\prod_{k=1}^N \langle{\uparrow_y}|_k ~ \sum_{i=1}^N ((\sigma_i)_x +
(\sigma_i)_y)
 ~ \prod_{j=1}^N |{\uparrow_x} \rangle_j}
\over  { \sqrt 2 ~ N(\langle{\uparrow_y} |{\uparrow_x} \rangle)^N}} = \sqrt 2
 ~~.
\eqno(14)
$$
The interaction Hamiltonian is
$$
H = {{g(t)}\over N} P \sum_{i=1}^N  (\sigma_i)_\xi~~~~.
\eqno(15)
$$
The initial state of the measuring device defines the precision of the
measurement. When we take it to be the Gaussian (3),  it is characterized by
the width $\Delta$. For a meaningful experiment we have to  take $\Delta$
small. Small $\Delta$  corresponds to large uncertain $P$,
 but now, the strength of the coupling to each individual spin is
reduced by the factor $1/N$. Therefore, for  large $N$,  both the
forward-evolving
state and the backward-evolving state are essentially not  changed by the
coupling to the measuring device. Thus,  this single measurement yields  the
weak value.  In Ref. 7 it is proven  that  if  the measured observable is an
average on a large set of systems,
$A = \bigl(\sum_i^N A_i \bigr)/N$, then we can  always construct a single,
good-precision measurement of the weak value.  Here let us present just
numerical calculations of
the probability distribution of the measuring device for $N$ pre- and
post-selected spin-1/2 particles. The state of the pointer after the
post-selection  for this case is
$$
{\cal N}\sum_{i=1}^N (-1)^i \bigl(\cos ^2 (\pi/8)\bigr)^{N-i} ~\bigl(\sin ^2
(\pi/8)\bigr)^i ~e^{-(Q-{{(2N-i)}\over N})^2/{2\Delta^2}} ~~. \eqno(16)
$$
The probability distribution for the pointer variable $Q$ is
$$
 prob (Q) ={\cal N}^2 \Bigl (\sum_{i=1}^N (-1)^i \bigl(\cos ^2
(\pi/8)\bigr)^{N-i} \bigl(\sin ^2
(\pi/8)\bigr)^i e^{-(Q-{{(2N-i)}\over N})^2/{2\Delta^2}}\Bigr)^2~~ .
 \eqno(17)
$$
The result for $N=20$ and different values of $\Delta$ are  presented in Fig.
4. We see that for $\Delta = 0.25$ and
larger, the obtained results are very good: the final probability
distribution of the pointer is peaked at the weak value, $\bigl((\sum_{i=1}^N
(\sigma_i)_\xi)/N\bigr)_w = \sqrt 2$. This distribution is
very close to that of a measuring device
measuring operator $O$ on a system in an eigenstate $|O{=} \sqrt2 \rangle$.
For $N$ large, the relative uncertainty can be  decreased  almost by a factor
$1/\sqrt N$ without
changing the fact that the peak of the distribution points   to
the weak value.

Although our  set of particles pre-selected in one state and post-selected in
another state is considered as one system, it looks very much as an ensemble.
 In quantum theory,
measurement of the sum does  not necessarily yield the same result as the
sum of  the results of the  separate measurements, so conceptually our
measurement on the set of particles  differs from the measurement on an
ensemble of  pre- and post-selected
particles. However, in our example of weak measurements, the results are
the same.

 Less ambiguous case is the
example considered in the first work on weak measure-ments.$^2$ In this
work a single system of a large spin $N$ is considered. The system is
pre-selected in the
state  $|\Psi_1\rangle =  |S_x {=} N \rangle$ and  post-selected in the
state  $|\Psi_2\rangle =  |S_y {=} N \rangle$.

\break

\phantom{xxxx}
\vfill

\noindent {\bf
Figure 4.~ Measurement on a single system.}~  Probability distribution of the
pointer variable for measurement of
$A =(\sum_{i=1}^{20}
(\sigma_i)_\xi)/20$  when the system of 20 spin-1/2 particles is
pre-selected in the state  $|\Psi_1\rangle = \prod_{i=1}^{20} |{\uparrow_x}
\rangle _i$ and post-selected in  the state
   $|\Psi_2\rangle = \prod_{i=1}^{20} |{\uparrow_y} \rangle _i$
.The  strength of the
measurement  is
parameterized by the width of the distribution $\Delta$. While in the very
strong
measurements, $\Delta = 0.01-0.05$, the peaks of the distribution located
at the eigenvalues, starting from $\Delta = 0.25$ there is essentially a
single peak at the location of the weak value, $A_w = \sqrt 2$.

\break
\noindent
 At an intermediate time the
spin component $S_\xi$ is weakly measured and  again the ``forbidden" value
$\sqrt 2 N$ is obtained. The uncertainty has to be only slightly larger than
$\sqrt N$.   The probability distribution of the
results is centered around $\sqrt 2 N$, and for large $N$ it
lies clearly outside the range of the eigenvalues, $(-N, N)$. Unruh$^{8}$ made
computer calculations of the distribution of the pointer variable for this
case and got results which are very similar to what is presented on Fig. 4.

An even more dramatic example is a measurement of the kinetic energy of a
tunneling
particle.$^{9}$ We consider a particle prepared in a bound state of a potential
well which has negative potential near the origin and vanishing potential
far from the origin; $|\Psi_1\rangle =  |E{=} E_0 \rangle$.  Shortly  later,
the particle is found
far from the well, inside a classically forbidden tunneling
region; this state can be characterized by vanishing potential
$|\Psi_2\rangle =  |U {=} 0 \rangle$. At an intermediate time a measurement of
the kinetic energy is
performed. The  weak value of the kinetic energy in this case is
$$
K_w =  {{\langle U{=}0 | K | E {=} E_0 \rangle}
\over{\langle U{=}0 | E{=}E_0 \rangle} } =
  {{\langle U{=}0 | E - U | E {=} E_0 \rangle}
\over{\langle U{=}0 | E{=}E_0 \rangle} } = E_0 ~~~~.
\eqno(18)
$$
The energy of the bound state, $E_0$, is negative, so the weak value of the
kinetic
energy is negative.  In order to obtain this negative value the coupling to
the measuring device   need not be too weak. In fact, for any finite
strength of the measurement we can choose the post-selected state
sufficiently far
 from the well to ensure the negative value. Therefore, for
appropriate post-selection, usual measurement of a
positive definite operator  invariably yields negative result!

 How do  we get this
paradoxical outcome? One can interpret it as a game of errors. Any
realistic experiment must have errors. Measurement of kinetic energy must
have a spread, so sometimes it might show negative outcomes. Of course, the
dial of the measuring device might have a pin preventing negative readings,
but we consider the device without such a pin. In our pre- and
post-selection measurement a peculiar interference effect of the pointer
takes place: destructive interference in the whole ``allowed" region and
constructive interference of the tails in the ``forbidden" negative region.
The initial state of the measuring device $\Phi (Q)$, due to the measuring
interaction and the post-selection, transforms into  a superposition of shifted
wave functions. The shifts are  by the
(possibly small) eigenvalues, but the  superposition is approximately
equal to the original wave function shifted by (large and/or forbidden)
weak value
$$
\sum_i  c_i \Phi (Q - a_i) \simeq  \Phi (Q - A_w)~~.\eqno(19)
$$
The example of a single weak measurement on the system of 20 pre- and
post-selected spin-1/2 particles  which was considered above  demonstrates
this  effect for a Gaussian wave function
of the measuring device, but we have proved$^7$ that ``miraculous" interference
(19) occurs not just for the  Gaussians, but  for a large class of functions.
The only requirement is that their
 Fourier transform must be   essentially bounded.

 It is possible to use this idea
for constructing a quantum time machine, a device which can
make  a cat out of a kitten in a minute.$^{7, 10}$  The superposition of
quantum states shifted by small periods of time can yield a large shift in
time; and it even can be a shift to the past.

These surprising, even paradoxical effects are really  gedanken
experiments. The reason is that, unlike weak measurements on an ensemble,
these are extremely rare events. For  yielding unusual weak value,  a
single  pre-selected system needs extremely improbable outcome of the
post-selection measurement.
Let us compare this with a weak  measurement on an ensemble. In order to get
$N$ particles in a pre- and post-selected ensemble which yield
$(\sigma_\xi)_w = 100$, we need $\sim N 10^4$ particles in the pre-selected
ensemble. But,  in order to get a single system of
$N$ particles yielding $(S_\xi)_w = 100 N$, we  need $\sim  10^{4N}$
systems of $N$ pre-selected particles. In fact, the probability to obtain
unusual value
by error is much larger than the probability to obtain the proper
post-selected state. We will see a negative reading of the device
measuring kinetic energy much faster than we will find the particle in  a deep
tunneling region. What makes these rare effects interesting is that there is
strong (although only one-way) correlation: every time we find in the
post-selection measurement the particle outside the well, we know that the
result of the
kinetic energy is negative, and not just negative: it is equal to the weak
value, $K_w = E_0$, with a good precision.

It is not that weak  measurement on a single pre- and
post-selected system cannot be measured in a laboratory. These are the
experiments with very dramatic results which are not feasible to perform.
 But an
experiment in which the weak value is only slightly outside the range of
the eigenvalues, performed on particles which  can be identically prepared  in
millions, is
possible.

Although we call it a weak measurement on  a single system, in practice the
experiment is performed on a large (pre-selected) ensemble. We
prepare many systems,  couple  each system to a  separate measuring device,
and make the post-selection measurement waiting for the desired
result. This is an experiment on a pre-selected ensemble, but it is an
experiment on a {\it single} pre- and post-selected system. Indeed, we discard
everything  connected to other systems. Only the reading of the measuring
device of  one system is considered. If we are ``lucky'' and the first
particle gets the right result in the post-selection measurement, then the
experiment is
completed, and only one system has been involved. This property does not
hold for  usual measurement on an ensemble: even if by chance the first result
yields  exactly the measured expectation value, we cannot stop here because we
cannot know yet that this is, indeed, the correct  value.

\vskip 1 true cm

\noindent
6. EXPERIMENTAL REALIZATIONS OF WEAK MEASUREMENTS
\vskip .5 true cm

Weak measurements have three basic elements: preparation, (weak) coupling
to the measuring device, and post-selection. The preparation part is the
same as in all usual experiments, so it does not require any special
consideration except that we need, for getting large effects, a very good
precision in the preparation of quantum state. The second stage too does
not present special experimental
difficulties: this is a  standard measuring procedure  with weakened
coupling. What limits the feasibility of a weak measurement  is the
possibility  of an effective  post-selection. In order
to obtain interesting results in weak measurement, the post-selection needs
to be very precise, and it has to fulfill a special requirement specified
below.

Realistic weak measurements (on an ensemble) involve preparation of a large
pre-selection ensemble, coupling to the
measuring devices of each element of the ensemble, post-selection
measurement which, in all interesting cases, selects only a small fraction
of the original ensemble, selection of corresponding measuring devices, and
statistical analysis of their outcomes. In order to obtain good precision,
this selected ensemble of the measuring devices has to be sufficiently large.
Although there are significant technological developments in ``marking''
 particles running in an experiment, clearly the most effective solution
is that the particles themselves serve as measuring devices. The
information about  measured variable is stored, after the weak measuring
interaction,  in their other degree of
freedom. In this case the post-selection of the particles in the required final
state automatically yield selection of measuring devices. The requirement
for the post-selection measurement is, then,  that there is no coupling
between  the
variable  in which the result of the weak measurement is stored and the
post-selection device.

 An example of such a case is the  Stern-Gerlach
experiment where the shift in the
momentum of a particle, translated into a spatial shift, yields the outcome of
the spin measurement. Post-selection measurement on a spin in a certain
direction can be implemented by another (this time strong) Stern-Gerlach
coupling which splits the beam of the particles. The beam corresponding to
the desired value of the spin is then analyzed for the result of the weak
measurement. The requirement of non-disturbance of the results of the weak
measurement by post-selection can be fulfilled by arranging the
 shifts due
to the two Stern-Gerlach devices to be orthogonal to each other. The
details are spelled out in Ref. 6.

A  weak measurement of a spin component is very
difficult to perform in a laboratory. We need very precise pre- and
post-selection of spin polarization and Stern-Gerlach experiment is very
far from being precise. But analogous experiments  can be performed on other
two-state systems. The simplest analog of the Stern-Gerlach measurement is
an  optical polarization experiment.
A birefringent prism splits an optical beam according to its polarization
modeling the inhomogeneous magnetic field which splits the beam of
particles with a spin. And high precision polarization filters serve as
excellent devices for  pre- and post-selection. We can define a polarization
operator
$$
Q \equiv |x\rangle \langle x| -  |y\rangle \langle y|  ~~, \eqno(20)
$$
where $ |x\rangle$ and  $|y\rangle$ designate photon linear polarization
states. The eigenstates of the polarization operator are $\pm 1$ but if the
initial state is $|\Psi_1\rangle = \cos \alpha ~ |x\rangle + \sin \alpha ~
|y\rangle$, and the final state is  $|\Psi_2\rangle = \cos \alpha ~ |x\rangle -
\sin \alpha ~
|y\rangle$, then the weak value of the polarization operator is
$$
Q_w = {{ (\cos \alpha \langle x| - \sin \alpha \langle y|)~
(|x\rangle \langle x| -  |y\rangle \langle y|)~
(\cos \alpha ~ |x\rangle + \sin \alpha ~ |x\rangle)}\over
{ \cos^2 \alpha  - \sin^2 \alpha}} = {1\over{\cos (2\alpha)}} ~~\eqno(21)
$$
 The initial and final states are chosen by placing
an appropriate linear polarization filters. If  the polarizations are  almost
orthogonal, $\alpha \simeq \pi/4$, the weak value of the polarization operator
becomes arbitrary
large.

 An analysis of realistic experiment which  can yield large weak value
$Q_w$ appears in Ref. 11. Duck, Stevenson, and Sudarshan$^{12}$ proposed
slightly different optical
realization which uses birefringent plate instead of a prism. In this case
the measured information is stored directly in the spatial shift of the
beam without being generated by the shift in the momentum.
 Ritchie, Story,   and Hulet adopted these scheme and performed the first
successful
experiment measuring weak value of the polarization operator.$^{13}$
 Their results are in very good agreement
with  theoretical predictions.  They obtained weak values which are very far
from the range of the eigenvalues, ($-1, 1$), their highest reported result
is $Q_w = 100$. The discrepancy between calculated and observed weak value
was   1\%. The RMS deviation from the mean of 16 trials was  4.7\%. The  width
of
the probability distribution was   $\Delta = 1000$ and  the number of pre- and
post-selected photons was $N \sim 10^{8}$, so the theoretical and experimental
uncertainties were of the same order of magnitude. Their other run, for
which they showed experimental data on graphs (which fitted very nicely
theoretical graphs), has the following characteristics:
$Q_w = 31.6$, discrepancy with calculated value 4\%, the RMS deviation 16\%,
$\Delta = 100$, $N \sim 10^5$.

Suter, M. Ernst and R. Ernst reported experimental realization of quantum
time-translation machine.$^{14}$ We may disagree about their experiment being a
model of our proposed time machine,$^{7,10}$ but it seems that they indeed
performed  a weak measurement.
The experiment was performed on $\, ^{13}C-\, ^1\! H$ spin pair of chloroform.
The
heteronuclear $J$ coupling of  the two spins $S$ and $I$ is given by
$$
H_{SI} = - 2\pi J S_z I_z ~~.\eqno(22)
$$
For a particular state of the spin $I$, the spin $S$ precesses due to this
spin-spin interaction.
In the experiment, an appropriate pre- and post-selection on the states of spin
$I$ were
performed, and it was observed that the spin $S$ precession was 4 times
faster than the one corresponding to the maximal eigenvalue of the second spin,
$I_z=1$. Rather than
 interpreting it as a time machine for $S$ we  see this experiment  as a weak
measurement of $I_z$, the measurement in which the weak value is 4 times
larger than the maximal eigenvalue, $(I_z)_w = 4$.

There are numerous experiments on pre- and post-selected systems. Post-
selection might lead to very dramatic effects  pointed out as
early as 1935 by Schr\"odinger.$^{15}$ Not all measurements on  pre- and
post-selected systems are weak measurements. Since some ``weak'' measurements
are not really weak, and some weak couplings are really  strong
measurements, it is not easy to find a rigorous definition for weak
measurements. A possible criterion is that
measurements yielding consistently weak values are weak measurements. Thus,
another run of the experiment of  Ritchie, Story,   and Hulet$^{13}$ which
shows dramatic effect is not a weak measurement. They considered
a post-selection to a state which is orthogonal to the initial state. The
probability for this post-selection was not zero due to the intermediate
weak  coupling. However, since  $|\Psi_2\rangle$ is orthogonal to
$|\Psi_1\rangle$,  the weak value is  not defined in this case.

Another system which is a good candidate  for weak measurements, due to a  well
developed technology of
preparation and selection of various quantum states, is a
Rydberg two-level atom. Between the pre- and post-selection the atom  can have
weak coupling with a
resonant field in a microwave cavity.$^{16,17}$

There are many  experiments measuring escape  time of tunneling
particles. Tunneling is a pre- and post-selection experiment: a particle
is pre-selected inside the bounding potential  and   post-selected
outside.
Recently, Steinberg$^{18}$ suggested that many of these experiments  are indeed
 weak measurements.

We believe that the field of experimental realization of weak measurements
is far from being exhausted. The next section explains the potential
applications of this procedure.

\vskip 1 true cm

\noindent
6.  WEAK MEASUREMENT AS AN AMPLIFICATION SCHEME
\vskip .5 true cm

Let us consider an experiment of a weak measurement of  $A$
not as a measurement of the weak value of $A$ but as a measurement of a
certain parameter of the measuring device. Indeed, when we consider known
initial
state $|\Psi_1\rangle$ and known final state $|\Psi_2\rangle$,
the weak value (1) is known prior to the measurement, and our experiment
yields no new information. But we can perform the weak measuring procedure
when the strength of the weak coupling is not known. Then, from the result of
the weak measurement we can find the strength of the coupling.

 The
Hamiltonian of the Stern-Gerlach experiment measuring $ z$ component of
a spin is
$$
H =- g(t) \mu {{\partial B_z}\over {\partial z}} z \sigma _z ~~. \eqno(23)
$$
We can prepare the state of the  spin, $ \sigma _z = 1$, then, assuming
that the gradient of the magnetic field is known,  our
experiment is a
von Neumann measurement of the magnetic moment $\mu$. (Or, if $\mu$ is known,
it
is  a measurement of the gradient.) Indeed, Eq. (23) has the form  of Eq.
(2) with $P$ replaced by $-z$ which corresponds to the pointer variable
$p_z$. The shift in the momentum is later transformed  into the shift in
the position of the particle.
 If we perform, instead, the pre- and
post-selected measurement with  the initial and final states of the spin
corresponding to, for example, $(\sigma_z)_w =100$, than our procedure is a
measurement of $\mu$ which is 100 times more sensitive! The shift in the
peak of the pointer position distribution is 100 times larger, while the
width of the peak is practically unchanged.

Of course, for increasing the  shift of the pointer we have to pay some price.
First, we cannot work with  narrow peaks. We have to be in a regime of weak
measurements, so the initial distribution of $z$ has to be well localized
around zero, and this requires wide pointer distribution ($p_z$). And
second, we lose the intensity. For obtaining amplification by factor $M$ we
need a post-selection which will reduce the number of systems of the
pre-selected ensemble by the factor of $1/M^2$, so in our example the
intensity will be reduced by the factor $10^{-4}$.

 Still, we believe that there
are measurements for which this amplification scheme might be  helpful. In
many experiments intensity is not a problem. If the output of the
measurement is a picture on a photographic plate, then the restriction is on
the total number of photons which were absorbed by the plate (before it was
saturated) while the number of pre-selected photons coming from a light
source is practically unlimited. This is the situation in the
optical analog of the Stern-Gerlach experiment discussed above. In optical
experiment, instead of
the magnetic moment  $\mu$, we measure  the degree of optical activity of the
crystal, i.e., the difference between indices of refraction for orthogonal
polarization, $n_x - n_y$. It seems that if one wants to measure this
difference using  a birefringent prism and the given light beam is suitable for
weak
measurements, then the post-selection certainly  increases the precision
of the measurement. If, however, our equipment allows us to make the incoming
beam well collimated, then the measurement without post-selection  has
an  advantage since it is  easier to find the center of a more narrow peak.
The analysis of an optimal measurement has not been performed
yet. But irrespective of the results of this theoretical analysis, we are
convinced that for realistic tasks when the  equipment is given, the scheme of
weak pre-
and post-selected measurement will prove itself useful for  improvement of the
sensitivity of
some measurements.

\vskip 1 true cm

\noindent
6. SUMMARY
\vskip .5 true cm

Weak measurement is a  certain  measuring procedure which includes
post-selec-tion. There are many peculiar effects due to various
post-selections, but weak measurements play  a special role  among them. The
outcomes of weak measurements,  weak values, are not just peculiar because they
are
very different from the outcomes of standard measurements: they are
part of new simple and rich structure existing in quantum world. The concept
of weak values is simple and universal. Weak values are  defined for all
variables and
 for all possible histories of quantum systems. They  manifest themselves  in
all
 couplings which are sufficiently weak.

The two basic elements of our approach were investigated separately.
The theory of ``unsharp'' measurements developed by Bush$^{19}$ has the element
of weakness of the interaction. Popular today, the ``consistent histories''
approach,$^{20}$ originated by Griffiths,$^{21}$ includes the idea of pre- and
post-selection. But it is the combination of the two which created our
formalism. It allowed us to see peculiar features of quantum systems. The
quantum time
machine, the method of increasing sensitivity  using post-selection, and
other surprising phenomena
were inaccessible within the framework of the  standard
formalism. Neither consistent histories nor unsharp measurements provided
tools to see these effects, although they might be helpful for  analyzing
these phenomena.$^{22}$

The formalism of weak measurement can also be helpful in describing
existing peculiar effects. The controversy of superluminal motion of
tunneling particles can be resolved by recognizing that the experiments showing
superluminal motion are weak measurements.$^{18}$ We have showed how, under
conditions of weak measurements, the
post-selection leads to superluminal motion of light wave packets (Sec. VIII of
Ref.~3).

Among applications of the weak value concept is a proposal to study
the back reaction of a quantum field on a particle-antiparticle pair
created by the field.$^{23}$  The weak value of the field is
considered
between the (initial) vacuum state and the (final) state which includes
the particle-antiparticle pair. The aim of this proposal is the
analysis of particle creation by a black hole and the problem of what
happens in the final stages of black hole evaporation.$^{24,25}$ One key to
this
problem is the back reaction of the pair to the gravitational field
that created it, and here the application of weak values signals the
possibility of a major breakthrough.

We have shown that the concepts of weak measurements and weak values are
useful tools. But one can speculate that it has more meaning than
this.$^{26,27}$
 The formalism of weak values is a candidate for describing reality of quantum
systems. It is well known that the usual concepts of reality developed from
centuries of thoughts based on classical physics fail to describe our world
which includes observed quantum phenomena.
It has been shown that there are severe difficulties in defining
relativistically invariant elements of reality in quantum theory. Weak values
are Lorentz invariant. So,  this might be the right course of action to
identify weak values  as elements of reality.$^{28}$ The weak values include
 the elements of reality as defined by Einstein, Podolsky and Rosen$^{29}${ as
well
as those recently defined by  Redhead.$^{30}$ In addition to universal
applicability,  weak values do have desirable
features such as the sum rule: if $C = A + B $ then $C_w = A_w + B_w$. However,
there is no product rule:  $C = A B$  does not lead to   $C_w = A_w
B_w$. But may be this is how our world really is.$^{31}$

\vskip 1.2 true cm

\noindent
 REFERENCES
\vskip .5 true cm

\noindent
1. Y.Aharonov, Bergmann,
and J.L.  Lebowitz, {\it Phys.  Rev. } {\bf B 134}, 1410 (1964).\hfil
\break
2. Y. Aharonov, D. Albert, A. Casher, and L. Vaidman,
 {\it Phys. Lett.} {\bf  A 124}, 199 \phantom{xxx}(1987).\hfil \break
3.  Y.Aharonov
and L. Vaidman, {\it Phys.  Rev.}   {\bf A 41}, 11 (1990); {\it J. Phys.} {\bf
A 24}, 2315
\phantom{xxx}(1991).\hfil \break
4. J. von Neumann,  `` Mathematical Foundations of Quantum Theory,''
  Princeton \phantom{xxx}University Press, New Jersey (1983). \hfil \break
5.  Y. Aharonov and L. Vaidman, this volume.\hfil \break
6. Y. Aharonov,  D. Albert, and L. Vaidman,
{\it Phys. Rev. Lett.} {\bf  60}, 1351 (1988).\hfil \break
7. Y. Aharonov, J. Anandan, S. Popescu, and L. Vaidman,
{\it Phys. Rev. Lett.} {\bf  64}, \phantom{xxx}2965 (1990).\hfil \break
8. W.G. Unruh, ``Time gravity and quantum mechanics," E-board:
gr-qc/9312027.\hfil

 \break \noindent
9. Y. Aharonov,  S. Popescu, D.~Rohrlich, and L.~Vaidman, {\it Phys.  Rev.}
{\bf A 48}, 4084 \phantom{xxx}(1993);
 {\it Jap. Jour. App. Phys.} {\bf 9}, 41 (1993).\hfil \break
10. L. Vaidman,
 {\it Found. Phys.} {\bf  21}, 947 (1991).\hfil \break
11. J.M. Knight and L. Vaidman
{\it Phys. Lett.} {\bf  A 143}, 357 (1990).\hfill \break
12. M.Duck, P.M. Stevenson, and E.C.G. Sudarshan, {\it Phys. Rev.} {\bf D 40},
2112 (1989).\hfil \break
13. N.W.M. Ritchie, J.G. Story, and R.G. Hulet, {\it Phys Rev. Lett.} {\bf
66}, 1107 (1991); \phantom{xxx}J.G. Story, N.W.M. Ritchie, and R.G. Hulet, {\it
Mod.
Phys. Lett.} {\bf B 5}, 1713 (1991).\hfill \break
14. D. Suter, M.Ernst, and R.R. Ernst, {\it Molec. Phys.} {\bf 78}, 93
(1990).\hfill \break
15. E. Schr\"odinger, {\it Proc. Camb. Phil. Soc.} {\bf 31}, 555 (1935).\hfill
\break
16. Y. Aharonov, L. Davidovich, and N. Zagury, {\it Phys. Rev.} {\bf A 48},1687
(1993).\hfill \break
17. N. Zagury and A.F.R. de Toledo  Piza, {\it   Phys. Rev.} {\bf A}, to be
published.\hfill \break
18. A.M. Steinberg, in {\it Fundamental Problems of Quantum Theory}, D.
Greenberger, \phantom{xxx}ed., NYAS (1994).\hfill \break
19. P. Bush,  {\it Phys. Rev.} {\bf D 33}, 2253 (1986).\hfill \break
20. M. Gell-Mann and J.B. Hartle, Phys. Rev. {\bf D 47}, 3345 (1993).\hfill
\break
21. R. B. Griffiths {\it J. Stat. Phys.} {\bf 70}, 2201 (1993).\hfill \break
22. P. Bush, {\it Phys. Lett.} {\bf  A 130}, 323 (1988).\hfill \break
23.$\!$ R. Brout, S. Massar, S. Popescu, R. Parentani, and Ph. Spindel,
 ``Quantum \phantom{xxx$\,$}source
of the back reaction on a classical field,"
E-board: hep-th/ 9311019 (1993).\hfill \break
24. F. Englert, S. Massar, R. Parentani, ``Source vacuum  fluctuations of
black hole \phantom{xxx}radiance,"  E-board: gr-qc/ 9404026 (1994).\hfill
\break
25. C.R. Stephens, G. 't Hooft and B.F. Whiting, ``Black hole
evaporation without \phantom{xxx}information loss,''  E-board:
gr-qc/9310006 (1993).\hfill \break
26. L. Vaidman, ``The problem of the interpretation of relativistic quantum
theories,'' \phantom{xxx}{\it Ph.D. Thesis}, Tel-Aviv University (1987).\hfill
\break
27. $\!$Y. Aharonov and D. Rohrlich, in {\it Quantum Coherence},  J.
Anandan, ed., World \phantom{xxx$\,$}Scientific, 221 (1990).\hfill \break
28. $\!\!\!$L. Vaidman,
{\it Symposium on the Foundations  of Modern Physics},
 P.J. Lahti, \phantom{xxx}P.~Bush, and P. Mittelstaedt, eds., World
Scientific, 406 (1993).\hfill \break
29. A. Einstein, B. Podolsky, and W. Rosen, {\it Phys. Rev.} {\bf 47}, 777
(1935).\hfill \break
30. M. Redhead, {\it Incompleteness, Nonlocality, and Realism} (Clarendon,
Oxford, 1987) \phantom{xxx}p. 72.\hfill \break
31. L. Vaidman, Phys.  Rev.  Lett. {\bf 70}, 3369 (1993).

\end